\documentclass[twocolumn, showpacs, preprintnumbers, amsmath, amssymb]{revtex4}
\usepackage{graphicx}
\usepackage{dcolumn}
\usepackage{bm}
\usepackage{color}
\begin{document}
\title{Gaussian Statistics of Fracture Surfaces}
\author{St{\'e}phane Santucci}
\affiliation{Department of Physics, University of Oslo, P.\ O.\ Box 1048
Blindern, N--0316 Oslo, Norway}

\author{Joachim Mathiesen}
\altaffiliation{Permanently at Department of Physics, University of Oslo, 
P.\ O.\ Box 1048 Blindern, N--0316 Oslo, Norway}
\affiliation{Department of Physics, Norwegian University of Science and
Technology, N--7491 Trondheim, Norway}

\author{Knut J{\o}rgen M{\aa}l{\o}y}
\email[K.J.Maloy@fys.uio.no]{} 
\affiliation{Department of Physics, University of Oslo, P.\ O.\ Box 1048
Blindern, N--0316 Oslo, Norway}

\author{Alex Hansen}
\affiliation{Department of Physics, Norwegian University of Science and
Technology, N--7491 Trondheim, Norway}

\author{Jean Schmittbuhl}
\affiliation{Institut de
Physique du Globe de Strasbourg, UMR CNRS 7516, 5, rue Ren{\'e}
Descartes, F--67084 Strasbourg, France}

\author{Loic Vanel}
\affiliation{Laboratoire de Physique, UMR CNRS 5672,  
Ecole Normale Sup{\'e}riure de Lyon,
46 All{\'e}e d'Italie, F--69364 Lyon, France}

\author{Arnaud Delaplace}
\altaffiliation{Permanently at LMT, Ecole Normale Sup{\'e}rieure de Cachan,
61, ave.\ du Pr{\'e}sident Wilson, F--94235 Cachan, France}
\affiliation{Department of Physics, University of Oslo, P.\ O.\ Box 1048
Blindern, N--0316 Oslo, Norway}

\author{Jan {\O}istein Haavig Bakke}
\affiliation{Department of Physics, Norwegian University of Science and
Technology, N--7491 Trondheim, Norway}

\author{Purusattam Ray}
\affiliation{Institute of Mathematical Sciences, Taramani, Chennai, 600 113,
India} 

\date{\today}
\begin{abstract}
We analyse the statistical distribution function for the height
fluctuations of brittle fracture surfaces using extensive 
experimental data sampled on widely different materials and
geometries. We compare a direct measurement of the distribution
to a new analysis based on the structure functions. For length
scales $\delta$ larger than a characteristic scale $\delta^*$, we find
that the distribution of the height increments $\Delta h = h(x+
\delta) -h(x)$ is Gaussian.  Self-affinity enters through the
scaling of the standard deviation $\sigma$, which is proportional to
$\delta^\zeta$ with a unique roughness exponent. Below the scale
$\delta^*$ we observe an effective multi-affine behavior of the height
fluctuations and a deviation from a Gaussian distribution which
is related to the discreteness of the measurement or of the
material.
\end{abstract}
\maketitle

It is difficult to believe that there may be anything in common
between the morphology of fractures in, say, concrete and aluminium,
except for the qualitative statement that they both are ``rough".  The
roughness seems very different when comparing the two
materials. Studies
\cite{mpp84-bs85,b97,mhhr92,blp90,blp90-sgr93-cw93-sss95-cpbfbgm03}
have shown that the scaling properties of this roughness are the same
to within the measuring accuracy for not only these two materials, but
for most brittle or weakly ductile materials that have been
tested. The scaling properties of the roughness alluded to above, is
more precisely described as the fractures being self-affine.  The
typical deviations $\Delta h$ of the surface as a function of distance
$\delta$ along the fracture surface scale as $\Delta h \propto
\delta^{\zeta}$
\cite{mpp84-bs85,b97,mhhr92,blp90,blp90-sgr93-cw93-sss95-cpbfbgm03}. It
has been suggested that these scaling properties might be {\it
universal\/} \cite{blp90,mhhr92}.

Most studies focus only on the scaling properties of the fracture
surfaces. They give no hint of the actual statistical distribution
giving rise to such a scaling.  In this study we go beyond a
calculation of the roughness exponents and propose a statistical
distribution for the height fluctuations $\Delta h$ of fracture
surfaces.  For the various materials and geometries we analyzed, we find
that the Gaussian distribution provides a complete statistical
description of the morphology of fractures at least at large scales.
This result is in contradiction with other works where multi-scaling 
\cite{bsv91}, more precisely, global multi-affinity is observed
\cite{bpssv05-bps05,sss95}.

Our work is based on the analysis of experimental data obtained
from various experiments on different materials.  The materials have
been broken in different modes and geometries and the surfaces have
been analyzed along one or more directions.

First we used the roughness measurement of a fracture surface obtained
from the failure of a granite block in mode I (4 bending point
failure). The scanned area of 10cm$\times$10cm covers the complete
section of the block with a grid mesh of
$(\delta_o)^2=$48$\mu$m$\times$48$\mu$m.  Accordingly the grid
size is: 2062$\times$2063, i.e.\ more than 4 million data points. The
profiler is optical with a laser beam of 30$\mu$m in
diameter \cite{strg04}. To reduce possible optical artifacts, the
scanned surface comes from a high-resolution silicon mold of the
granite fracture. The replica technique in a perfectly homogeneous
material removes fluctuations of local optical properties and
significantly improves the quality of the roughness measurement.

Second, we used data from interfacial fracture fronts propagating (in
mode I) into the annealing plane of two plexiglas (PMMA) plates
\cite{smts06,sm97-sdmpv2003,dsm99,msts05}.  We have analyzed 6 long
front lines (obtained by assembling fronts for a crack at rest
\cite{dsm99}) containing 17000 pixels each, with a pixel size
$\delta_o=2.6 \mu$m.  The roughness exponent was found equal to
$\zeta=0.63 \pm 0.03$ \cite{dsm99}.  

Finally, we studied fracture surfaces obtained during fracture
experiments on a quasi two-dimensional material: fax paper sheets
loaded in mode I at a constant force \cite{svc04,scdvc05}. High
resolution scans were performed on post-mortem samples.  We analyzed 5
fronts with around 10000 pixels each, the pixel size is $\delta_o= 20
\mu$m.  In that case the roughness exponent measured is $\zeta^{2d}
\approx 0.6$.

We aim at estimating the statistical distribution function of
the height fluctuations of a fracture surface $P (h(x+\delta)-h(x) )$.
However, a direct measurement of the distribution function is not
always accessible due to limited statistics. Only the first data
set which is very large, will allow us to perform such a direct
estimate. For the others, we propose a method \cite{h91} introduced in
connection with the study of directed polymers \cite{hh85-k85-k85b}
that is based on the structure functions defined as the $k^{th}$ root
of the $k^{th}$ moment of the increment $ \vert \Delta h \vert = \vert
h(x+\delta) -h(x) \vert $ on a scale $\delta$ :
\begin{equation}
\label{structfunc}
C_k(\delta)=\langle \vert h(x+\delta)-h(x)\vert^k\rangle^{1/k}\; .
\end{equation}
The average is taken over the spatial coordinate $x$.  Now, forming
the ratio between the $k^{th}$ structure function and the second
structure function, we define the function
\begin{equation}
R_k (\delta) = {{\langle \vert h(x+\delta)-h(x)\vert^k\rangle^{1/k}} \over 
{\langle (h(x+\delta) - h(x))^2\rangle^{1/2}}}\; .
\label{eqRatios}
\end{equation}

\begin{figure}
\includegraphics[width=7cm,clip]{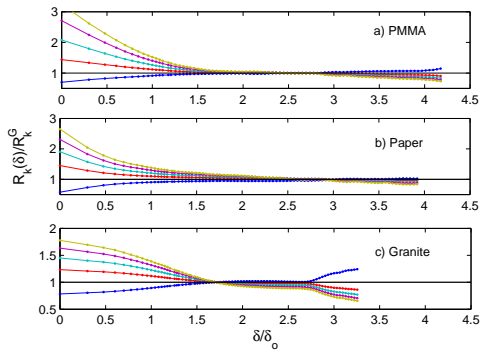}
\caption{Convergence of the moment ratios $R_k$ as function of
$\delta$ towards the Gaussian ratios $R_k^G$ for the different 
order $k=1,2,...,6$ , confirming the 
underlying Gaussian statistical distribution. 
For each set, on the small scales, the individual 
lines represents from below to above increasing $k$ values. 
The various ratios $R_k$ are 
averaged over a) $6$ interfacial fronts in PMMA, b) $5$ fracture 
fronts in Paper 
and $100$ profiles c) perpendicular to the 
fracture propagation in a granite block.} 
\label{fig1}
\end{figure}

In the case of a one-dimensional Brownian motion, the statistical
distribution of increments $\Delta h$ is a Gaussian:
\begin{equation}
P( \Delta h )={1\over{\sqrt{2 \pi \sigma^2}}}\ e^{-(\Delta h)^2/2\sigma^2}\; ,
\label{eq4}
\end{equation}
Note that the self-affinity enters through the variance of the
distribution $ \sigma^2 \propto \delta^{2\zeta}$, where $\zeta = 1/2$
for the Brownian motion.  Below we find the same statistical
distribution for the various fracture fronts investigated except that
generally $\zeta \neq 1/2$.  For the Gaussian distribution the moments
Eq.\ (\ref{structfunc}) are easily calculated, $C_k^G(\delta)=(2
\delta^{2\zeta})^{1/2}\left(\Gamma ((k+1)/2)/\sqrt{\pi}\right)^{1/k}$.
In this case the ratios $R^G_k$ of the structure functions become
\begin{equation}
R^G_k =\sqrt{2} \left(
{{\Gamma \left( {{k+1} \over 2} \right)} 
\over {\sqrt{\pi}}}\right)^{1/k}\; .
\label{eqGaussRatios}
\end{equation}
Note that these Gaussian ratios are independent of $\sigma^2$ and
$\delta$; they contain no adjustable parameters.
The generality of this result transcends the
derivation based on an underlying Gaussian distribution. A different
underlying distribution will give rise to a different set of ratios
$R_k$.

First, we computed the moment ratios normalized by the Gaussian
values, $R_k(\delta)/R_k^G$, and in particular their variation with
$\delta$ for the different fracture profiles we have studied. In Fig.\
\ref{fig1} we show that the ratios $R_k(\delta)$ converge for
different values of $k$ and for large $\delta$, towards the values of
a Gaussian process, suggesting the Gaussian nature of the distribution
for a range of $\delta$ values. Note that the deviations from a
constant value for the largest values of $\delta$ happen when the
statistics become very poor.

\begin{figure}
\includegraphics[width=7cm,clip]{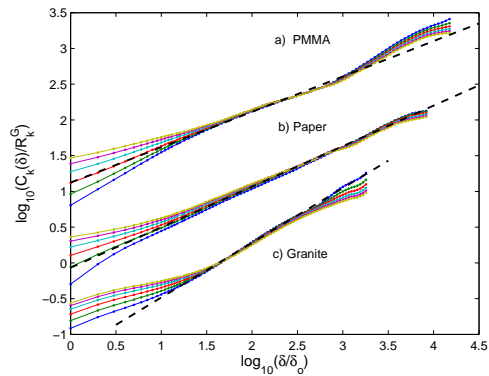}
\caption{ Data collapse of the structure function normalized by the
Gaussian ratios $C_k(\delta)/ R_k^G$ for the values $k=1,2,...,6$. The
various data set are displaced vertically to improve the visual
clarity.  The dashed lines are fit to the second order structure
functions $C_2(\delta)/ R_2^G$ on a range where the structure
functions collapse. Their slopes provide an estimate of the roughness
exponent. We estimate the following exponents, for PMMA $\zeta \approx
0.5$, for paper $\zeta^{2d} \approx 0.6$ and for granite
$\zeta^{3d}\approx 0.8$, respectively.}
\label{3d}
\end{figure}
Second, we examined the scaling behavior of the structure
functions. In Fig.\ \ref{3d} we show directly that the structure
functions collapse, when normalized by the Gaussian ratios,
$C_k(\delta)/R_k^G$.  The collapse at the larger scales provides clear
evidence that no multi-scaling is present, i.e.\ the scaling exponent of
$C_k(\delta)\sim \delta ^{\zeta_k}$ is independent of $k$ and
therefore we may extract a unique roughness exponent $\zeta_k=\zeta$.
A fit to the second order structure functions $C_2(\delta)/ R_2^G$ on
the range where the data collapse provides an estimate of $\zeta$
within 10\%. For the granite surface we find a roughness exponent
around $\zeta^{3d}\approx 0.8$. For the interfacial fracture fronts in
PMMA, we obtain a roughness exponent $\zeta \approx 0.5$, 
which is slightly lower than but consistent with previous estimates
\cite{sm97-sdmpv2003,dsm99,msts05,smts06}. Finally, for paper we
observe an exponent $\zeta^{2d}\approx 0.6$.

\begin{figure}
\includegraphics[width=8cm,clip]{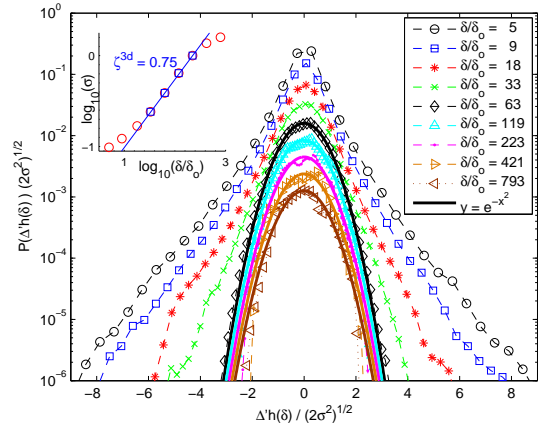}
\caption{Statistical distributions of the height fluctuations
$P(\Delta' h)$ sampled from a grid of $2000$ lines in the direction
perpendicular to the fracture propagation in a 
3D granite block. We show the distribution for
logarithmically increasing length scales 
$\delta$. Note that in addition, we have
shifted the various distributions 
logarithmically for visual clarity. We plot
on a semilog scale $P(\Delta' h)\sqrt{2 \pi \sigma^2}$ versus
$(\Delta' h)/ \sqrt{2 \sigma^2}$ and observe at large scales a typical
parabolic shape of a Gaussian distribution. The solid lines represent
the curve $y=e^{-x^2}$ and fit perfectly the experimental
distributions above a characteristic 
length scale $\delta/\delta_o>\delta^\star
\sim 50 $. Inset: The scaling behavior of the standard
deviation of the distributions $P(\Delta' h(\delta))$
allows us to extract the roughness of the fracture surface: $\sigma
\propto \delta^{\zeta^{3d}}$ with $\zeta^{3d} \approx 0.75$.} 
\label{Distri}
 \end{figure}

The fact that the rescaling by Gaussian ratios leads to a data
collapse, suggests that the underlying distribution is Gaussian.  This
result is confirmed by a direct analysis of the large data set from 
the fracture surface in a granite block.  The analyzed data set
consisted of $2000 \times 2000$ points representing the central
part of a 3D map of the fracture surface to reduce boundary
effects.  From the map we not only computed the structure
functions as shown in Fig.\ \ref{3d}, but we also computed directly the
statistical distribution of the height fluctuations $P(\Delta' h)$ at
different length scales $\delta$, where $\Delta' h = (\Delta h
-\langle \Delta h \rangle)(\delta)$.  Note that we subtract the
averaged height fluctuations $\langle \Delta h \rangle$ in order to
center the various distributions around a zero mean. The structure
functions (Eq.\ (\ref{structfunc})) are defined without such a
procedure.  However, we checked that it did not influence the scaling
behavior of the structure functions, by directly detrending the
various profiles.  In Fig.\ \ref{Distri} we show the distributions of
the height fluctuations for logarithmically increasing length scales
$\delta$. The data were extracted in the direction perpendicular to
the fracture propagation and the distributions were sampled from the
$2000$ profiles $h(x)$ each containing $2000$ points.  We clearly see
that above a characteristic length scale $\delta^\star \sim 50 \times
48 \mu$m the shape of the distributions become Gaussian.  This scale
corresponds to the point of convergence observed in Fig.\ \ref{fig1}
and the onset of collapse in Fig.\ \ref{3d}.  Interestingly, a similar
cross over has been observed in the width distribution of contact
lines measured recently \cite{mrkr04}. We emphasize that the
self-affine behavior of the fracture front enters through the scaling
of the standard deviation $\sigma \propto \delta^{\zeta^{3d}}$ with
$\zeta^{3d} \approx 0.75$, see the inset in Fig.\ \ref{Distri}.
 
Finally, the length-scale independence of Eq. (\ref{eqGaussRatios})
and the data collapse in Fig. \ref{3d} provide clear evidence that
there is no multi-scaling of the structure functions. That being
said, we do observe a separation of the structure functions at small
scales together with a broadening of the tails of the distributions
observed in Fig. \ref{Distri}. Several authors have reported similar
findings \cite{bbjkvz92,bmr05-llr05}. In KPZ models on kinetic surface
roughening a broadly distributed noise gives rise to rare but large
perturbations of the surface and hence a multi-scaling of the structure
functions at small scales \cite{bbjkvz92}.

To illuminate the multi-scaling at small scales, consider a piecewise
continuous fracture surface with a number of vertical jumps of size
$\epsilon_i$, e.g.\ due to overhangs, microscopical defects or the
grain size in granite and the fibre size in paper.  On sufficiently
small length scales $\delta$, the height variations in the surface
will be negligible relative to the jump size $\epsilon_i$. The
structure function will therefore collect from each jump a
contribution roughly proportional to $\delta\epsilon_i^k$
\cite{m02}. Overall, the contribution to Eq.\ (\ref{structfunc}) will
be $C_k(\delta) \sim \delta^{1/k}(\sum_i \epsilon_i^k)^{1/k}$ where
the sum is taken over all the jumps. We see that the structure
function now scales with a $k$-dependent Hurst exponent
$\zeta_k=1/k$. This behavior is observed for values of $k$ close to or
larger than unity. For small values of $k$ the effect of the vertical
jumps will diminish relative to the ordinary surface roughening and
therefore $\zeta_k\approx \zeta$ for $k\ll 1$.

\begin{figure}
\includegraphics[width=8cm,clip]{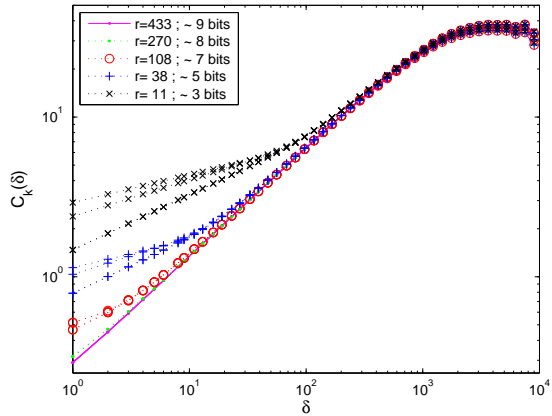}
\caption{Influence of the coarse graining at small scales on the
structure function $C_k(\delta)$, (k=2,3,4), for synthetic fractional 
Brownian motions with a roughness exponent $\zeta=0.7$. The 100 samples
have been coarse grained using different resolutions corresponding to
represnting $h$ by 3, 5, 7, 8 or 9 bits.  This corresponds to using from 
$r=11$ to 433 distinct values for spanning the range of $h$-values.  The 
grouping of the data sets into three different classes correspond to the 
three values of $k$, with $k=4$ at the top.  Note that the separation of the 
structure functions on the small scales diminishes as the resolution is 
increased.}
\label{noise}
\end{figure}

A few comments should be given on this $k$-dependence. First of all,
it is questionable to base arguments in favor of multi-affinity on the
smallest scales. It is important to have in mind that the experimental
scan of the surface at small scales might be spurious due to
limitations of the profilometers or scanners and contain artificial
jumps due to overhangs etc. The discretization (coarse graining) of the
data also plays a crucial role, see \cite{bprr00}.  
Fig.\ \ref{noise} demonstrates
how sensitive the structure function is to the discretization. We
generated graphs with roughness exponent $\zeta=0.7$ from a
fractional Brownian motion. We then filtered the graphs by
representing the values of $h$ using 3, 5, 7, 8 and 9 bits (see Fig.\
\ref{noise}). When decreasing the resolution (the number of bits), we
observe at small scales a clear deviation from the expected scaling
behavior and more importantly a separation of the structure functions
$C_k(\delta)$. The grain size in granite and the fiber size in the
paper experiment may introduce a discreteness similar to that of
sampling the fracture surface at a low resolution. For example, in the
granite experiment, we observe a crossover from a microscopic
multi-scaling to a pure self-affine surface around the typical grain
size (order of 1 mm), see Fig.\ \ref{Distri}.  The multi-scaling, or more
precisely, the separation of the structure functions at small scales,
thus, most likely reflects the experimental limitations or the
microscopic details of the material.

In conclusion, we have analyzed experimental data on fracture profiles
in widely different materials and have shown that the structure
function ratios $R_k$ (Eq. \ref{eqGaussRatios}) converge to the values
of a Gaussian process. We have verified our findings by also computing
directly the distribution of the height fluctuations and have shown
that there exists a scaling exponent, $\zeta$, permitting the
rescaling of the height fluctuation distribution $P(\Delta h(\delta))
\sim \delta^{-\zeta}G(\Delta h( \delta)/\delta^\zeta)$.  We
find that the rescaling function $G$ has a Gaussian form. From a
fundamental point of view, the distribution of the height fluctuations
provides new important information about the morphology of fracture
surfaces; information which is not covered by the calculations
of a roughness exponent. In addition, the wide-ranging convergence of
the ratios $R_k$ shows that there is no global multi-scaling of the
structure functions.

We thank S.G.\ Roux and R.\ Toussaint
for fruitful discussions and their critical reading of the manuscript.  
S.\ Santucci was supported by the NFR Petromax program 163472/S30, J.\
Mathiesen was supported by NFR-166802, J.\ Schmittbuhl by the EHDRA
project.



\begin{thebibliography}{99}

\bibitem{b97} 
E.\ Bouchaud, J.\ Phys.\ Condens.\ Matt.\ {\bf 9}, 4319 (1997).

\bibitem{mpp84-bs85} 
B.\ B.\ Mandelbrot, D.\ E.\ Passoja, and A.\ J.\
Paullay, Nature, {\bf 308}, 721 (1984); S.\ R.\ Brown and C.\ H.\
Scholz, J.\ Geophys.\ Res.\ {\bf 90}, 12575 (1985).

\bibitem{blp90} 
E.\ Bouchaud, G.\ Lapasset, and J.\ Plan{\'e}s, Europhys.\ Lett.\ {\bf 13}, 
73 (1990).

\bibitem{mhhr92} 
K.\ J.\ M{\aa}l{\o}y, A.\ Hansen, E.\ L.\ Hinrichsen and S.\ 
Roux, Phys.\ Rev.\ Lett.\ {\bf 68}, 213 (1992).

\bibitem{blp90-sgr93-cw93-sss95-cpbfbgm03} 
J.\ Schmittbuhl, S.\ Gentier, and
S.\ Roux, Geophys.\ Res.\ Lett.\ {\bf 20}, 639 (1990); B.\ L.\ Cox and
J.\ S.\ Y.\ Wang, Fractals, {\bf 1}, 87 (1993); F.\ C{\'e}lari{\'e}, 
S.\ Prades, D.\ Bonamy, L.\ Ferrero, E.\ Bouchaud, C.\ Guillot, and 
C.\ Marli{\`e}re, Phys.\ Rev.\ Lett.\ {\bf 90}, 075504 (2003).

\bibitem{bsv91}
A.\ L.\ Barab\'asi, P.\ Sz\'epfalusy and T.\ Vicsek, Physica A, {\bf 178},
17 (1991).

\bibitem{bpssv05-bps05} 
E.\ Bouchbinder, I.\ Procaccia, S.\ Santucci, L.\ Vanel, Phys.\ Rev.\ Lett.\ 
{\bf 96}, 055509 (2006).

\bibitem{sss95}
J.\ Schmittbuhl, F.\ Schmitt and C.\ Scholtz, J.\ Geophys.\ Res.\
{\bf 100}, 5953 (1995).

\bibitem{strg04} 
J.\ Schmittbuhl, R.\ Toussaint, F.\ Renard, and J.\ P.\ Gratier, Phys.\ 
Rev.\ Lett.\ {\bf 93}, 238501 (2004).

\bibitem{dsm99} 
A.\ Delaplace, J.\ Schmittbuhl, and K.\ J.\ M{\aa}l{\o}y, 
Phys.\ Rev.\ E, {\bf 60}, 1337 (1999).

\bibitem{msts05} 
K.\ J.\ M{\aa}l{\o}y, S.\ Santucci, R.\ Toussaint and J.\ Schmittbuhl, 
Phys.\ Rev.\ Lett.\ {\bf 96}, 045501 (2006).

\bibitem{smts06} 
S.\ Santucci, K.\ J.\ M{\aa}l{\o}y, R.\ Toussaint and
J.\ Schmittbuhl, in {\it Dynamics of Complex Interconnected Biosensor 
Systems: Networks and Bioprocesses,\/} A.\ T.\ Skjeltorp ed.\ (Kluwer, 
Amsterdam, 2006).

\bibitem{sm97-sdmpv2003} 
J.\ Schmittbuhl and K.\ J.\ M{\aa}l{\o}y, Phys.\ Rev.\ Lett.\ {\bf 78},
3888 (1997); J.\ Schmittbuhl, A.\ Delaplace, K.\
J.\  M{\aa}l{\o}y, H.\ Perfittini and J. P. Vilotte, Pageoph, {\bf 160},
961 (2003).

\bibitem{svc04}
S.\ Santucci, L.\ Vanel and S.\ Ciliberto, Phys.\ Rev.\ Lett.\ {\bf 93},
095505 (2004).

\bibitem{scdvc05} 
S.\ Santucci, P.\ Cortet, S.\ Deschanel, L.\ Vanel and S.\ Ciliberto, 
Europhys.\ Lett\ {\bf 74}, 595 (2006).

\bibitem{h91} 
T.\ Halpin-Healy, Phys.\ Rev.\ A, {\bf 44}, R3415 (1991).

\bibitem{hh85-k85-k85b} 
D.\ A.\ Huse and C.\ L.\ Henley, Phys.\ Rev.\ Lett.\ {\bf 54}, 2708 (1985); 
M.\ Kardar, Phys.\ Rev.\ Lett.\ {\bf 55}, 2235 (1985); 
{\it ibid.\/} {\bf 55}, 2923 (1985).. 

\bibitem{mrkr04} S.\ Moulinet, A.\ Rosso, W.\ Krauth, and E.\ Rolley, 
Phys.\ Rev.\ E, {\bf 69}, 035103 (2004).

\bibitem{bbjkvz92} 
A.\ L.\  Barab\'asi, R.\ Bourbonnais, M.\ H.\ 
Jensen, J.\ Kert\'esz, T.\ Vicsek, and Y.\ C.\ Zhang, Phys.\ Rev.\ A, {\bf 45},
R6951 (1992).

\bibitem{bmr05-llr05} 
G.\ M.\ Buend\'ia, S.\ J.\ Mitchell and P.\ A.\ Rikvold, 
Microelectronics J.\ {\bf 36}, 913 (2005).

\bibitem{m02} 
S.\ J.\ Mitchell, Phys.\ Rev.\ E, {\bf 72}, 065103 (2005).

\bibitem{bprr00}
J.\ Buceta, J.\ Pastor, M.\ A.\ Rubio and F.\ J.\ de la Rubia,
Phys.\ Rev.\ E, {\bf 61}, 6015 (2000).

\end{thebibliography}
\end{document}